\documentclass[aip,apl,reprint]{revtex4-1}
\usepackage{graphicx}

\usepackage{sidecap}

\begin{document}
\title{Compact silicon double and triple dots realized with only two gates.}
\author{M. Pierre}
\author{R. Wacquez}
\author{B. Roche}
\author{X. Jehl}
\author{M. Sanquer}
\email[]{marc.sanquer@cea.fr}
\affiliation{CEA-DSM-INAC-SPSMS, 38054 Grenoble, France}
\author{M. Vinet}
\affiliation{CEA/LETI-MINATEC, 38054 Grenoble, France}
\author{E. Prati}
\author{M. Belli}
\affiliation{MDM-INFM, 20041 Agrate-Brianza, Italy}
\author{M. Fanciulli}
\affiliation{MDM-INFM, 20041 Agrate-Brianza, Italy}
\affiliation{Dipartimento di Scienza dei Materiali, Universita degli Studi Milano-Bicocca, 20125 Milano, Italy}

\date{\today}

\begin{abstract}
We report electronic transport on silicon double and triple dots created with the optimized number of two gates. Using silicon nitride spacers two dots in series are created below two top gates overlapping a silicon  nanowire. Coupling between dots is controlled by gate voltages.  A third dot is created either by combined action of  gate voltages or local doping depending on the spacers length. The main characteristics of the triple dot stability diagram are quantitatively fitted.
\end{abstract}
\maketitle

Semiconducting quantum dots are usually made with three gates: one plunger gate to control the density of carriers and two gates to control the in and out tunneling rates.
A recent achievement is a controlled silicon Single Electron Transistor (SET) using only one gate. This MOS-SET departs from the usual MOSFET \cite{Hofheinz06B,Hofheinz06} by a specific design of channel junctions to source and drain. Tunnel barriers are formed by undoped silicon segments and a quantum dot is created by accumulation of carriers below the gate. The undoped silicon barriers sit below spacers which are used as a mask during source and drain implantation. 

In this work we show that the association of two MOS-SETs in series --- the most natural extension of the single MOS-SET --- yields both a well controlled double dot with tunable inter-coupling and a triple dot. Within other technologies triple dots usually require many gates. Triple dots have been realized in 2DEG \cite{Waugh95,Schroer07,Vidan04} and in silicon using top and side gates \cite{Lee00}.
The advantage of a single gate design becomes crucial when compact arrangements of multiple quantum dots are required, for instance
to create high fidelity electron pumps,  latching switches, quantum cellular automata \cite{Orlov01}, single-electron-parametron devices \cite{Korotkov98}, non-local interacting qubits \cite{Hollenberg06,Greentree04} or rectifier devices \cite{Stopa02,Vidan04,Ferrari05}.

Recently  silicon quantum dot molecules, i.e.\ a double dot with tunable coupling, have been built  with one larger upper gate plus four side gates \cite{Shin07}
or three top gates \cite{Liu08}. Reducing further the number of gates has been possible at the cost of stochasticity: a coupled quantum dot was obtained with two top and one upper gates by relying on the static disordered potential arising from the Si/SiO$_2$ buried oxide interface rugosity \cite{Liu08B}. One advantage of our technology is that our double and triple dots are smaller and/or simpler than previously reported \cite{Waugh95,Schroer07,Vidan04,Lee00,Shin07,Liu08}.

\begin{figure}[!t]
\includegraphics[width=1 \columnwidth]{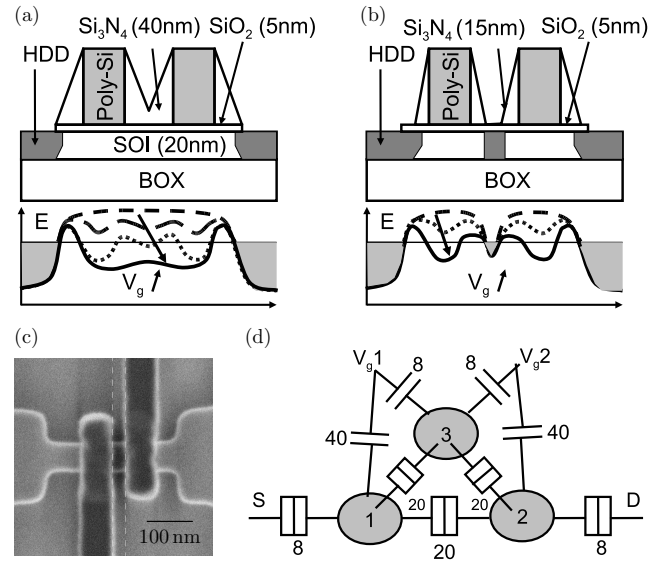}
\caption{schematic layout of the coupled MOS-SETs: a) long silicon nitride spacers protecting the central region of SOI during source-drain implantation (samples 1 and 2).
b) with short spacers a dot is created by implanted arsenic donors in the central region (sample 3). Below each case the schematic profiles of the bottom of the conduction band are drawn for various gate voltages. The horizontal line is the Fermi energy fixed by the source and drain. c) SEM micrograph of a typical sample before spacers deposition. The gate length is 60\,nm and the spacing between gate 30\,nm. d) Equivalent circuit for the triple dot system. The capacitance values corresponding to sample 2 are given in aF.}
\label{fig1}
\end{figure}

\begin{SCfigure*}[][t]
\includegraphics[width=1.4\columnwidth]{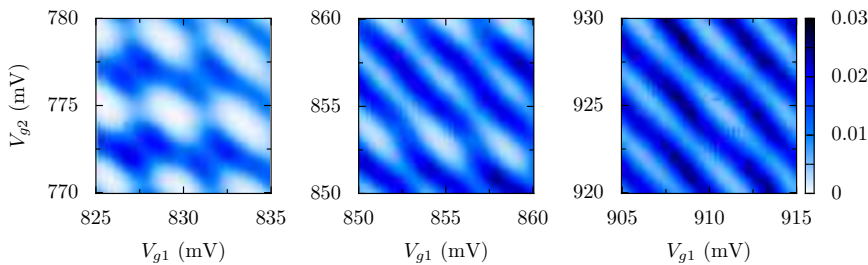}
\caption{ Source-drain conductance at $T=1$\,K versus gate voltages for sample 1: $W=60$\,nm, $L_g=60$\,nm, $L_{gg}=40$\,nm and 40\,nm spacers. From left (two capacitively coupled MOS-SETs) to right (single dot) gate-controlled interdot coupling increases.}
\label{fig2}
\end{SCfigure*}

 Our samples are produced on 200\,mm silicon-on-insulator (SOI) wafers in a CMOS platform. A 200\,nm long, 20\,nm thick and $W=20$--60\,nm wide silicon nanowire is etched and covered by two $L_g = 30$--60\,nm long polysilicon gates isolated by 5\,nm thick SiO$_\mathrm{2}$ gate oxide (Fig.~\ref{fig1}).
The spacing between gates is $L_{gg} = 30$--60\,nm. 15 or 40\,nm thick silicon nitride spacers are deposited on both sides of the gates. Therefore for samples with long (40\,nm) spacers (samples 1 and 2) the silicon region between the gates is undoped as it is masked during the heavy As implantation of the source and drain. On the contrary it is doped for samples with short (15\,nm) spacers (sample 3). Samples with thinner SOI (10\,nm) have also been studied and the increase of the period of Coulomb blockade oscillations (CBO) shows that the size of each MOS-SET is further reduced (not shown). The typical charging energy of our MOS-SETs is 2\,meV.

Figure~\ref{fig2} shows the drain-source current versus gate voltages applied on each gate at $T=1$\,K in a device with 40\,nm long spacers. Two nominally identical MOS-SETs appear below the two gates, with CBO periods very constant in gate voltage \cite{Hofheinz06B}.
Similar CBO periods and threshold voltages are observed for the two MOS-SETs, illustrating the reproductibility of the fabrication process.
At low ($V_{g1} , V_{g2}$) the two dots are decoupled and the 2D plot exhibits the expected square lattice pattern \cite{Vanderwiel03} (not shown).
At intermediate ($V_{g1} , V_{g2}$) the dots are capacitively coupled and the 2D plot shows the so-called honeycomb pattern (Fig.~\ref{fig2}, left panel). The gate capacitances $ C_{g1}$, $ C_{g2}$ and the interdot capacitance $ C_{12}$ are estimated to 40\,aF, 40\,aF and 20\,aF respectively.
At large ($V_{g1} ,  V_{g2}$) the tunnel coupling becomes non negligible and the 2D plot exhibits characteristic wandering anti-diagonals \cite{Vanderwiel03,Liu08} (Fig.~\ref{fig2}, central panel). At even larger gate voltages the two dots merge into a single island (Fig.~\ref{fig2} right panel). In that case the period along the diagonal in ($V_{g1} , V_{g2}$) plot is half that of the two separated dots, as expected since the gate capacitance is approximately twice larger for the single merged dot.
This evolution, similar to the ones reported in Ref. \onlinecite{Liu08,Liu08B}, shows that a tunable double dot is obtained with only two gates. The absence of  independent control on the carrier's number and interdot coupling is compensated by the compacity
and simplicity of our devices. 

\begin{SCfigure*}[][b!]
\includegraphics[width=1.4 \columnwidth]{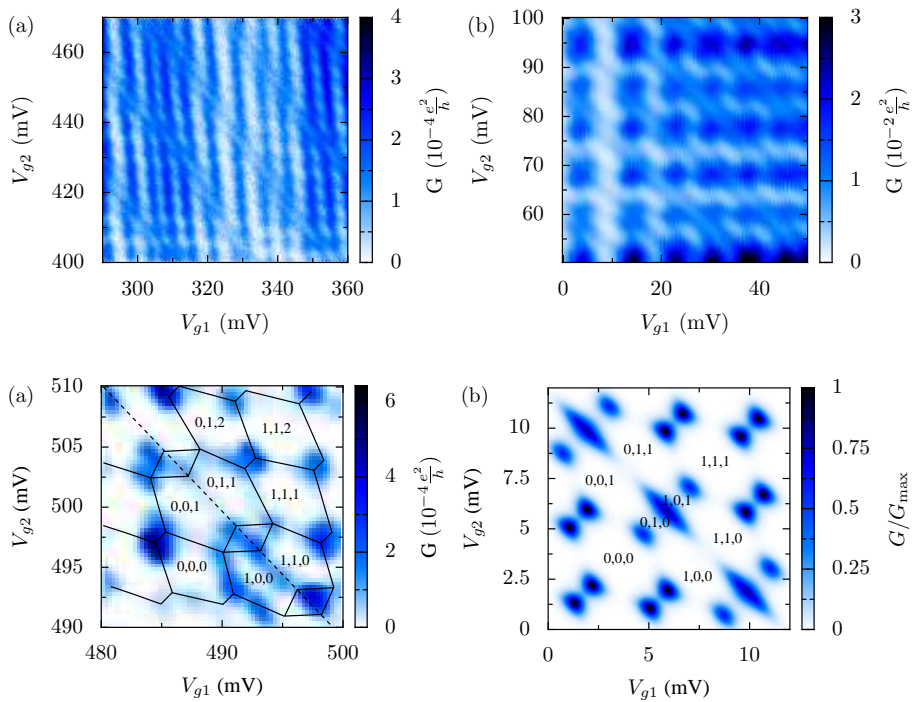}
\caption{Source-drain conductance versus gate voltages at $T=4.2$\,K. a) sample 2: $W=60$\,nm, $L_g=50$\,nm, $L_{gg}=50$\,nm and 40\,nm spacers. b) Sample 3: $W=60$\,nm, $L_g=60$\,nm, $L_{gg}=40$\,nm and 15\,nm spacers. Periodic anti-diagonals occur at each degeneracy point of a central dot formed between the two gates. The period is 2.5 larger for sample 2 (large spacers) compared to sample 3 (small spacers), indicating that the central dot is bigger due to local doping. In that case the source-drain current appears at lower gate voltages.}
\label{fig3}
\end{SCfigure*}

The transition from two independent dots towards a single island is realized by increasing simultaneously the two gate voltages which both control the potential of the undoped silicon nanowire segment separating the two MOS-SETs. This evolution is not monotonic. A periodic antidiagonal (i.e.\ constant $V_{g1}+ V_{g2}$ value) pattern  is superimposed (see Fig.~\ref{fig3}).
This period results from Coulomb blockade on a third intermediate dot created in the silicon nanowire, at equal distance from both gates. It arises from the combined action of the two unscreened gate potentials in samples with long spacers (see Fig.~\ref{fig1}a). The whole system then behaves as three dots in series controlled by two gate voltages. In order to model this triple dot we consider the electrostatic configuration  depicted in Fig.~\ref{fig1}d. The conductance is simulated within the orthodox model of Coulomb blockade without cotunneling \cite{Hofheinz06,Pierre09}. Cotunneling through dot 3 is nevertheless taken into account as a direct tunnel coupling between dots 1 and 2.
For the sake of simplicity the five tunnel junctions tunneling rates are taken equal. The capacitance values indicated in Fig.~\ref{fig1}d give the best fit for the periods measured in sample 2. These values also describe well sample 1. We note that these tunnel capacitances are larger than their simple planar geometrical approximations ($C_{12} =20$\,aF instead of 3\,aF for instance). This effect is attributed to the large polarizability of electrons in the tunnel barriers with small energy height \cite{Hofheinz07}.

\begin{SCfigure*}
\includegraphics[width=1.4 \columnwidth]{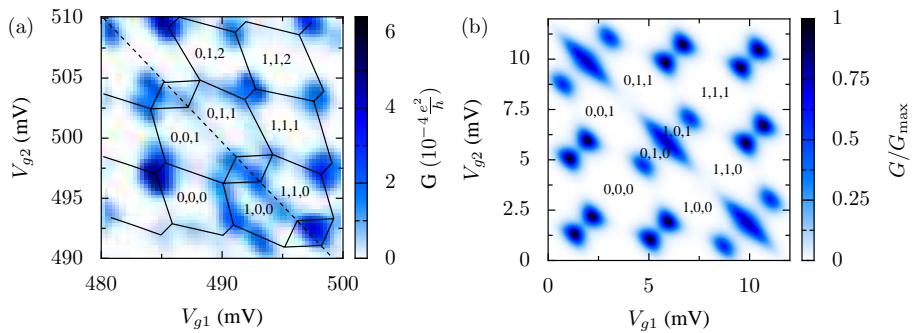}
\caption{ a) Source-drain conductance versus gate voltages for sample 2 at $T=1$\,K near degeneracy of the central dot. Hexagonal cells are distorted into pentagonal and diamond-shaped features. b) simulation at $T=1$\,K of the triple dot   sketched in Fig.~\ref{fig1}d). ($N_{1}$,$N_{3}$,$N_{2}$) are indicated with an arbitrary origin 
((0,0,0) corresponds to about (60,20,60)) because the figure is periodic in $N_{1}$,$N_{2}$,$N_{3}$ .} 
\label{fig4}
\end{SCfigure*}

The tunnel coupling between the two MOS-SETs as well as the occurence of the third dot can also be tuned by doping. Indeed in samples with small spacers (sample 3) the central region of the SOI nanowire is not protected and is therefore n-doped (see  Fig.~\ref{fig1}b). Thus at zero gate voltages the central region contains electrons.
This doped architecture is reminiscent of the locally implanted silicon quantum dots \cite{Jamieson05}.
The voltage at which source-drain current is detected at low temperature is  $V_{g1} = V_{g2}\simeq -0.1V$ in short spacers samples, whereas it is detected at larger values ($+0.2$\,V) in long spacers samples.
The honeycomb pattern for both kind of samples exhibit similar periods, showing that the characteristics of each MOS-SET is not affected by the central doping.
On the contrary the central dot is strongly modified: the period  of the anti-diagonals is reduced by a factor 2.5, which indicates a bigger central dot more capacitively coupled to gates (see Fig.~\ref{fig3}b).

On Fig.~\ref{fig4}a a detail of the diagram is shown at lower temperature ($T=1$\,K) in sample 2 for one of these anti-diagonals. The corresponding simulation shown in Fig.~\ref{fig4}b reproduces the distorted honeycomb pattern experimentally observed, except for lines joining the triple points which are a consequence of cotunneling. The number of electrons ($N_{1},N_{3},N_{2}$) on each dot is indicated in Fig.~\ref{fig4}b.
 $N_{1} \simeq 60$, $N_{2} \simeq 60$, $N_{3}\simeq 20$ are determined by dividing $V_{g}$ counted from the threshold ( $+0.2$\,V) by the CBO period and the figure is periodic
in $N_{1}$,$N_{2}$,$N_{3}$ (see Fig.~\ref{fig3}). Far from the degeneracy ($N_{3} \leftrightarrow N_{3}+1$) of dot 3 one recovers the honeycomb pattern for the double dot system. Near degeneracy we observe an alternance of pentagones and diamonds. Up to eight triple points per cell are predicted for a triple dot \cite{Schroer07}. In case of high symmetry and perfect matching between the capacitances these eight points can reduce to four quadruple points \cite{Schroer07}, as observed in both our simulation and data (see Fig.~\ref{fig4}).

In conclusion we have shown that the MOS-SET device \cite{Hofheinz06B} can be extended to multiple dots with a minimal number of gates. A triple dot has been realized for the first time in silicon with only two control gates. It consists of two MOS-SETs in series separated by a locally n-doped or undoped silicon nanowire.
The obtained triple dot is very compact and can be used to design more complex circuits based on single electron transfers.

The research leading to these results has received funding from the European Community's seventh Framework (FP7 2007/2013) under
the Grant Agreement Nr:214989. The samples subject of this work have been designed and made by the AFSID Project Partners http://www.afsid.eu.
E.P., M.B. and M.F. would like to acknowledge S. Cocco (Laboratorio Nazionale MDM) who contributed to develop the the transport acquisition equipment.


\begin{thebibliography}{19}

\bibitem{Hofheinz06B} M. Hofheinz, X. Jehl, M. Sanquer, G. Molas, M. Vinet and S. Deleonibus, Appl. Phys. Lett. {\bfseries 89}, 143504 (2006).
\bibitem{Hofheinz06} M. Hofheinz, X. Jehl, M. Sanquer, G. Molas, M. Vinet and S. Deleonibus, Eur. Phys. J. B {\bfseries 54}, 299 (2006).
\bibitem{Waugh95} F. R. Waugh, M. J. Berry, D. J. Mar, R. M. Westervelt, K. L. Campman and A. C. Gossard, Phys. Rev. Lett. {\bfseries 75}, 705 (1995).
\bibitem{Schroer07} D. Schr\"{o}er, A. D. Greentree, L. Gaudreau, K. Eberl, L. C. L. Hollenberg, J. P. Kotthaus and S. Ludwig, Phys. Rev. B {\bfseries  76}, 075306 (2007).
\bibitem{Vidan04} A. Vidan, R. M. Westervelt, M. Stopa, M. Hanson and A. C. Gossard, Appl. Phys. Lett. {\bfseries 85}, 3602 (2004).
\bibitem{Lee00} S. D. Lee, K. S. Park, J. W. Park, J. B. Choi, S.-R. E. Yang, K.-H. Yoo, J. Kim, S. I. Park and K. T. Kim, Phys. Rev. B {\bfseries 62}, R7735 (2000).
\bibitem{Orlov01}A. O. Orlov, R. K. Kummamuru, R. Ramasubramaniam, G. Toth, C. S. Lent, G. H. Bernstein and G. L. Snider, Appl. Phys. Lett. {\bfseries 78}, 1625 (2001).
\bibitem{Korotkov98} A. N. Korotkov and K. K. Likharev, J. of Appl. Phys.  {\bfseries 84}, 6114 (1998).
\bibitem{Hollenberg06} L. C. L. Hollenberg, A. D. Greentree, A. G. Fowler and C. J. Wellard, Phys. Rev. B {\bfseries 74}, 045311 (2006).
\bibitem{Greentree04} A. D. Greentree, J. H. Cole, A. R. Hamilton and L. C. L. Hollenberg, Phys. Rev. B {\bfseries  70}, 235317 (2004).
\bibitem{Stopa02} M. Stopa, Phys. Rev. Lett. {\bfseries  88}, 146802 (2002).
\bibitem{Ferrari05} G. Ferrari, L. Fumagalli, M. Sampietro, E. Prati, M. Fanciulli,  J. of Appl. Phys. {\bfseries 98}, 044505 (2005).
\bibitem{Shin07} S. J. Shin, J. J. Lee, R. S. Chung, M. S. Kim, E. S. Park,  J. B. Choi, N. S. Kim, K. H. Park, S. D. Lee, N. Kim and J. H. Kim, Appl. Phys. Lett.  {\bfseries 91}, 053114 (2007).
\bibitem{Liu08} H. Liu, T. Fujisawa, H. Inokawa, Y. Ono, A. Fujiwara and Y. Hirayama, Appl. Phys. Lett.  {\bfseries 92}, 222104 (2008).
\bibitem{Liu08B}H. W. Liu, T. Fujisawa, Y. Ono, H. Inokawa, A. Fujiwara, K. Takashina and Y. Hirayama, Phys. Rev. B {\bfseries  77}, 073310 (2008).
\bibitem {Vanderwiel03} W. G. Van der Wiel, S. De Franceschi, J. M. Elzerman, T. Fujisawa, S. Tarucha and L. P. Kouwenhoven, Review Modern Phys. {\bfseries 75}, 1 (2003).
\bibitem{Pierre09} M. Pierre, M. Hofheinz, X. Jehl, M. Sanquer, G. Molas, M. Vinet and S. Deleonibus, Eur. Phys. J. {\bfseries 70}, 475 (2009).
\bibitem{Hofheinz07} M. Hofheinz, X. Jehl, M. Sanquer, G. Molas, M. Vinet and S. Deleonibus, Phys. Rev. B {\bfseries 75}, 235301 (2007).
\bibitem{Jamieson05} D. N. Jamieson, C. Yang, T. Hopf, S. M. Hearne, C. I. Pakes, S. Prawer, M. Mitic, E. Gauja, S. E. Andresen, F. E. Hudson, A. S. Dzurak and R. G. Clark, Appl. Phys. Lett. {\bfseries 86}, 202101 (2005).

\end{thebibliography}
\end{document}